\documentclass[slac_one]{revtex4}
\usepackage{graphicx}
\usepackage{fancyhdr}
\newcommand{\upmu}{\mu}

\newcommand{\keV}{\,\mathrm{ke\kern-1pt V}}
\newcommand{\MeV}{\,\mathrm{Me\kern-1pt V}}
\newcommand{\GeV}{\,\mathrm{Ge\kern-1pt V}}

\newcommand{\TeV}{\,\mathrm{Te\kern-1pt V}}
\newcommand{\TEV}{\mathrm{Te\kern-1pt V}}

\newcommand{\mm}{\,\mathrm{mm}}

\newcommand{\mum}{\,\upmu\mathrm{m}}

\newcommand{\m}{\,\mathrm{m}}

\newcommand{\pb}{\,\mathrm{pb}}

\begin{document}

\title{{\small{Hadron Collider Physics Symposium (HCP2008),
Galena, Illinois, USA}}\\ 
\vspace{12pt}
Commissioning of the ATLAS Experiment} 

%
%

\author{Juergen Thomas on behalf of the ATLAS Collaboration}
\affiliation{University of Birmingham, Birmingham B15 2TT, United Kingdom,
juergen.thomas@cern.ch}

\begin{abstract}
The status of the commissioning of the ATLAS experiment as of May 2008 is 
presented. The sub-detector integration in recent milestone weeks is 
described. Cosmic commissioning in milestone week M6 included 
simultaneous data-taking and combined track analysis of the muon detector and 
inner detector, as well as combined analysis of muon detector and muon trigger.
The calorimeters have achieved 
near-full operation, and are integrated with the calorimeter trigger. The 
high-level-trigger infrastructure is being installed and algorithms tested in 
technical runs. 
\end{abstract}

\maketitle

\thispagestyle{fancy}

\section{INTRODUCTION}

The ATLAS collaboration consists of around 1900 scientific authors, from 165 
institutes in 35 countries. The detector is roughly cylinder-shaped with 
a height of 46$\m$ and 25$\m$ in diameter. It is installed in a cavern 92$\m$ below 
ground at CERN. A 'ship-in-a-bottle' assembly has been performed, as the 
cavern is just large enough for the detector. This document focusses on 
progress of commissioning the detector since the report at HCP2007 \cite{Amelung}.
%
%
%
\section{ATLAS DETECTOR COMPONENTS}

\begin{figure*}[t]
\centering
\includegraphics[width=110mm]{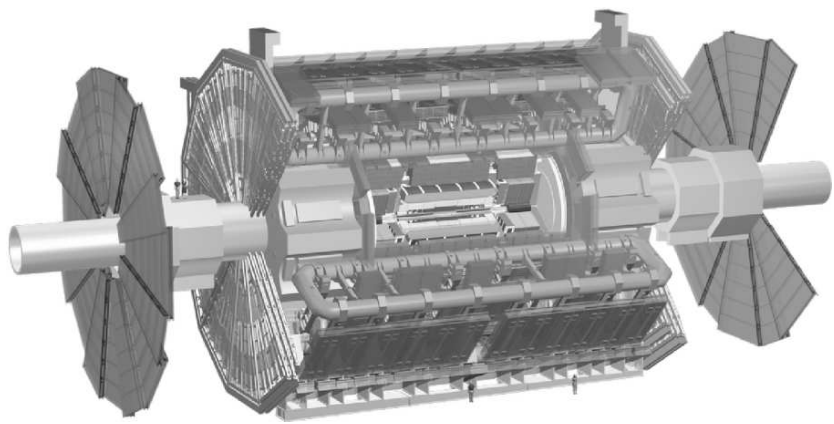}
\caption{The ATLAS detector. From inside to outside: Inner detector with
Pixel, SCT and TRT, then calorimeters (Liquid argon and scintillating tiles),
and muon spectrometer outside with toroid magnets.} \label{figure:1}
\end{figure*}

The detector's scale and architecture is determined by the requirements of 
the physics goals of the LHC programme, most notably excellent energy resolution of 
the calorimeters is required, as well as very good muon momentum resolution, 
and inner detector performance for heavy flavour identification.
A detailed description of the ATLAS detector is given in \cite{TechProposal}. 
An overview is given below:

\begin{itemize}
\item {\bf Inner Detector} with silicon pixel detector (Pixel) closest to the beam-line,
silicon strips detector (SCT) and transition radiation tracker (TRT), all
located inside a 2~T solenoidal magnetic field. These detectors provide 
precision tracking of charged particles and secondary vertex finding in the pseudo-rapidity 
region $|\eta| <$ 2.5. 

\item {\bf Calorimeters}: Liquid argon calorimeter (LAr) for 
electromagnetic barrel and endcap calorimeters, and hadronic endcaps (HEC) and 
forward calorimeters (FCAL). The electromagnetic calorimeter 
provides coverage for $|\eta| <$ 3.2, the limit 
of hermiticity is $|\eta| =$ 4.9. Steel and scintillator tiles are used
in the barrel region of the hadronic calorimeter (Tile). 
%

\item {\bf Muon spectrometer}:   
consisting of monitored drift tubes (MDT) in the barrel and endcap regions 
for precision measurements, and cathode strip chambers (CSC) in the forward region, 
integrated into the air core toroid magnet system. Embedded are the fast-responding 
muon trigger chambers, which are Resistive Plate Chambers (RPC) 
in the barrel and thin gap chambers (TGC) in the forward region.
The main requirement is precise muon momentum measurements within $|\eta| <$ 2.7.

\item {\bf Trigger and Data Acquisition} architecture: Three-level data 
selection architecture. The first level (Level-1) in custom-build 
parallel-processing 
pipelined hardware is separated into calorimeter and muon Level-1 trigger.
The high level triggers (HLT) consisting of the
second level (Level-2) and third level (Event Filter) are
implemented in software running on large PC farms with dedicated network
infrastructure. 
The event rate of 40 MHz from the detector is reduced to 200 Hz for event-data recording.\\
Readout is performed using custom-built buffers in 
PCs (Readout System: ROS). Data acquisition (DAQ) software to 
control, configure and monitor all systems.
\cite{TechProposal}.

\end{itemize}

\section{WORKING TOWARDS DATA-TAKING}

During the last year, the focus of the commissioning effort has evolved 
from single detector operation to combined running and integration. 
Monthly integration weeks are scheduled 
to integrate detector, trigger and data acquisition into one
global setup for each group of sub-detectors (calorimeters, muon detectors and
inner detector), which are then combined together for the milestone weeks.
All experts are brought together for those weeks, of which the sixth one, M6, 
took place in April, and M7 in May 2008. 
%
%
%
%
In addition, technical runs 
are performed, feeding simulated and recorded data into the data acquisition 
system to perform full-rate tests at Level-1 trigger rates of up to 50 KHz.
The start-up schedule as of May 2008 expects the ATLAS detector to be 
closed by mid-July, with first LHC injections by the end of July, and high-energy 
proton-proton-collisions at 10 TeV by September 2008. A few weeks of 
stable running is planned, producing a few $\pb^{-1}$ of data.
The M6 milestone week included all detector parts apart from the Pixel 
detector, with sizable percentages of each sub-detector included into the 
data acquisition, especially almost the entire complement of 
barrel calorimeters, muon barrel 
system and the Level-1 trigger systems for both.

Cosmic muons are very useful for the detector commissioning. They do 
however occur with very low rate and also do not originate in the 
interaction region to really mimic LHC events \cite{Amelung}. Different 
strategies are followed to make the detection of cosmic muons as useful as 
possible, modifying the standard data acquisition and reconstruction 
tools, for example by increasing the recorded time-interval and 
using calorimeter 
data to form track-like objects. Other important commissioning tools are the 
calibration pulser systems built into the detector front-end electronics, 
and a laser calibration system for the scintillating tile calorimeter.

\section{CALORIMETER COMMISSIONING}

The Liquid Argon calorimeter endcaps were fully switched on during the April 
Calo Week after the M6 run. Cosmic muons were used to run in a specially extended 
32-sample recording mode and also in 5-sample physics mode.
This mode records 32 samples of 25~ns each, unlike the 5-sample-mode in normal
LHC operation, which will be the identified bunch-crossing itself plus the two
leading and two trailing samples, and in addition with filter coefficients applied
in the calorimeters.
Timing studies with the trigger systems, both the first stage (Level-1) and the high
level triggers (HLT), have been performed, as well as monitoring and data quality tools 
being tested and improved. 
The full Liquid Argon calorimeter system was operational and was read out-during the M7 
week. Events triggered by the Level-1 calorimeter trigger were studied, and 
good timing alignment was achieved for the whole detector.  
Comparison with simulated pulse shapes showed good agreement. 
The Level-1 calorimeter trigger algorithms foreseen for LHC operations were 
set-up and adjusted, and trigger objects and results were analysed 
and verified using samples recorded in combined data-taking.
Cosmic muon tracks and pulse shapes were studied using event display and monitoring tools.

\begin{figure*}[t]
\centering
\includegraphics[width=120mm]{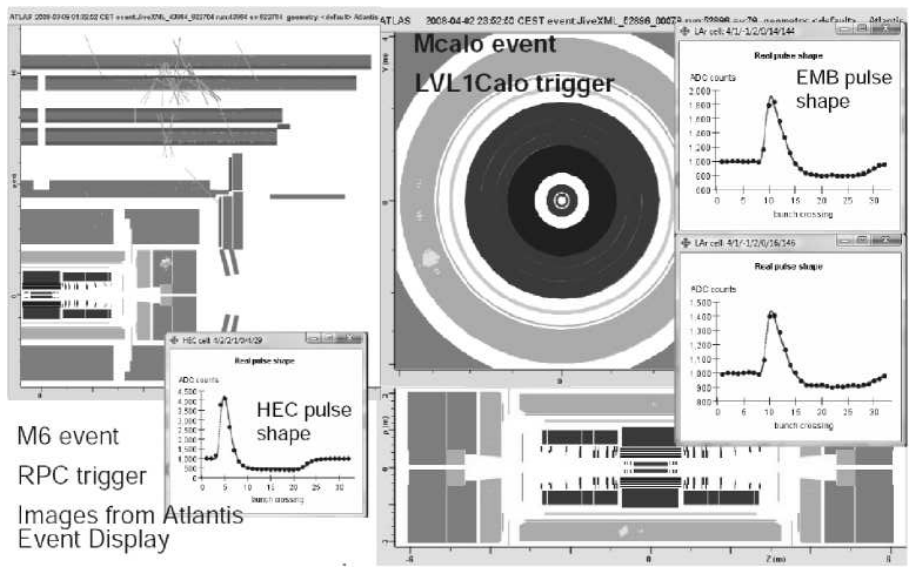}
\caption{LAr Calorimeter ADC counts and fitted pulse-shapes 
hadronic endcap (HEC) and electromagnetic barrel (EMB) systems
from commissioning in M6 Milestone week
 and energy deposition in event display (April 2008).} \label{figure:7}
\end{figure*}

The hadronic barrel calorimeter, consisting of scintillating tiles, 
was operational at 95\%  during M6 with the remaining modules undergoing power supply 
refurbishment. 
The detector uses a specially developed algorithm ('TileMuonFitter') for 
commissioning, which forms track-like objects from the calorimeter cell data, 
as they would be expected from a cosmic muon passing through. This
shows a good energy density peak. No top-bottom bias 
has been seen in the detector. A laser calibration system is commissioned 
to send light into the photomultiplier tubes (PMTs) to align the timing, 
and its operation integrated into the 
data acquisition. The pulse-shapes from the ADC counts are inspected and 
compared with the Level-1 trigger pulse-shapes. Fig. \ref{figure:7} shows
the pulses as analog-to-digital converter
(ADC) counts and fitted pulse-shapes for both the LAr hadronic endcap (HEC) 
and the electromagnetic barrel LAr calorimeter (EMB), along with 
an event display of the energy depositions.

\section{MUON DETECTOR COMMISSIONING}

\begin{figure*}[t]
\centering
   \begin{minipage}{145mm}
      \includegraphics[width=60mm, angle=0]{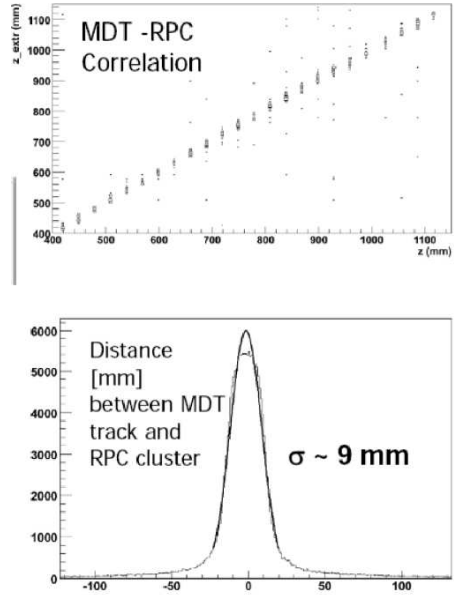}
   \end{minipage}
\caption{ 
Correlation of muon spectrometer (MDT) and embedded
Level-1 muon trigger chambers (RPC) from commissioning with cosmic muons
(February 2008). {\bf Top:} Correlation of extrapolated z coordinate from
both systems in $\mm$, {\bf bottom:} Distance between clusters in $\mm$.
}\label{figure:muon}
\end{figure*}

The barrel and endcap muon spectrometer (MDT) consists of 16 sectors, 
out of which 12 were ready to operate in the milestone week M6.
%
%
The upper sectors have already been commissioned with cosmic muons. 
This effort continues with the remaining four sectors in the lower half. 
%
%
An example of cosmic data analysis is given in Fig. \ref{figure:muon}, where
clusters from the muon trigger chambers (RPC) are compared with tracks in the 
precision muon spectrometer chambers 
(MDT), with a very good correlation observed. The residual distribution
shows a width of 9~mm. The MDT system
shows very good track quality for cosmic muons, with six hits per track, the 
residuals centred at zero, and a spread in the distribution (RMS) of 160$\mum$.
Timing calibration of 
the resistive plate chambers (RPC) as part of the Level-1 muon trigger system 
is performed, where the 
trigger settings are aligned between the planes using offsets from the 
time-of-flight distribution. Monitoring and event display tools have been
used to achieve synchronised read-out and investigate hits counts, noise levels 
and track quality.

\section{INNER DETECTOR COMMISSIONING}

\begin{figure*}[t]
\centering
\includegraphics[width=120mm]{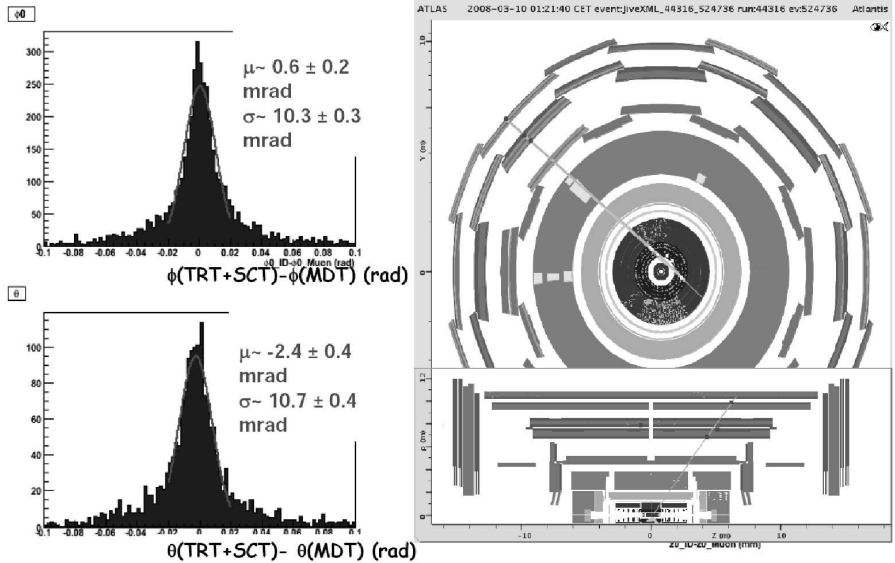}
\caption{{\bf Left:} Combined analysis of track parameters $\phi$ and $\Theta$ 
for inner detector (SCT and TRT) and muon spectrometer (MDT),
using data from M6 Milestone week
(April 2008), {\bf Right:} Event display of a reconstructed cosmic muon track 
seen in both the inner detector and muon spectrometer.} \label{figure:indet}
\end{figure*}

The inner tracking detector commissioning efforts have been seriously 
disrupted by the break-down of the cooling compressor at the beginning of May 
after only 5 days of 
Pixel detector commissioning in-situ. The compressors are being repaired. 
The SCT has not participated in later milestone weeks as it shares its cooling
infrastructure with the Pixel detector.
The Pixel detector has not been in combined cosmics running due to the problems
with the cooling system, and also as its commissioning was queued behind the 
SCT commissioning. 
%
%
%
Combined performance studies of the SCT and TRT have been performed using
cosmic data taken during earlier milestone weeks.
Alignment and calibration studies and improvements reduce the 
spread of the residual distribution considerably, e.g. for the TRT from 
450$\mum$ to 270$\mum$. In a further step, the track position
as measured by the inner detector (SCT and TRT) and in addition by 
the precision muon spectrometer (MDT) 
were combined, this is shown in Fig. \ref{figure:indet}, along with an event
display of a cosmic muon track seen in all three sub-detectors.
This uses the top-half muon spectrometer available in this milestone week.
For the angle $\phi$, the correlation width of $\sigma$ = 10.3
mrad is achieved, while for angle $\Theta$ the width is $\sigma$ = 10.7 mrad.
Such an analysis requires all the systems to be operational, timed-in and 
read-out successfully.

\section{TRIGGER AND DATA ACQUISITION}

The trigger and data acquisition architecture 
is described in detail in \cite{TechProposal}. The Level-1 
calorimeter trigger signals need to be thoroughly tested before access to 
the detector ends. Level-1 muon trigger commissioning is done sector by 
sector as they are connected to their respective gas and power supplies. 
The timing is being 
addressed to ensure all signals used for the trigger decision are 
synchronous, throughout the whole system. 
Data analysis shows that the hits and clusters recorded in the trigger system are 
well correlated with those reconstructed from the detector read-out.
%
The high-level trigger (HLT) computing farms are being commissioned and tested, 
at a rate of about ninety rack mounted PCs per week, and they perform second and third
stage trigger algorithm (Level-2 and Event Filter) tasks during the 
commissioning runs. 
Five server-level PCs, with sufficient buffer diskspace to hold
many hours of data, form the last stage of the on-site data acquisition. 
Then the data is transfered via a dedicated link to the CERN computing center, 
where it is reconstructed and made available for analysis.
%
%
The track trigger implemented into the second trigger stage (Level-2) 
was studied in cosmic data from the M6 milestone week, 
regarding its efficiency for events with high-momentum reconstructed 
tracks and its selectiveness of events to go into different output streams.
%
%
Dedicated TDAQ technical runs use simulated events, as well as recorded cosmic 
data from the earlier M4 and M5 weeks, at the expected rate of data at LHC
collisions.
%
%
The high-level trigger infrastructure and algorithms are commissioned 
to correctly identify cosmic-ray tracks when compared to the reconstructed
quantities. These studies use a transparent trigger mode, where events are
flagged with high-level trigger results, but not discarded accordingly,
as it would happen in standard operation.
In the technical runs, 
stable operation was achieved for hours without intervention, with the system 
controlling 1500 applications in 350 nodes.
%
%

\section{THE FUTURE}

Looking ahead, the activities during beam commissioning in single-beam 
operation will include validation of the beam protection systems, first 
synchronisation with the LHC clock and detailed timing and alignment 
studies, and feedback to the machine. With collisions, the trigger 
systems will be fully synchronised with the LHC. Full understanding of the 
whole detector has to be achieved, using well-known physics processes 
\cite{TatjanaTalk}.

\section{SUMMARY}
ATLAS is in the process of commissioning the detector using cosmic rays and calibration
systems. The sizeable fraction of sub-detectors is already integrated with
the trigger and data acquisition systems, allowing for stable combined data-taking
and combined data analysis. 
An intense commissioning programme still lies ahead to bring 
the components so far not integrated into the combined running into operations
and achieve stable data taking with the full detector
in time for the LHC start-up.
%
%
%
%
\begin{acknowledgments}
The results presented here are the result of the work of many ATLAS 
colleagues, and their contributions are gratefully acknowledged. Material 
has been provided by the ATLAS Commissioning Working Group chaired by 
Jamie Boyd and Maria Costa, by Ludovico Pontecorvo, Jose Maneira, Pippa 
Wells, Stephen Hillier and more. I'd like to thank the organisers of the 
conference for providing an inspiring atmosphere in a beautiful location. 
\\
This work is supported by the Science and Technology Facilities 
Council (STFC) in the UK.
\end{acknowledgments}

\end{document}